\documentclass[preprint,10pt]{aastex}

\usepackage{amssymb}
\usepackage{graphicx}
\usepackage{natbib}

\newcommand{\fn}[1]{\footnote{\scriptsize{#1}}} 

\begin{document} 

\slugcomment{\textit{Geophys. Res. Lett.} \textbf{37}, L14205 (2010), doi:10.1029/2010GL043663.}
\title{Cassini imaging search rules out rings around Rhea} 

\author{Matthew S. Tiscareno$^1$, Joseph A. Burns$^{1,2}$, Jeffrey N. Cuzzi$^3$, and Matthew M. Hedman$^1$}
\affil{$^1$Department of Astronomy, Cornell University, Ithaca, NY 14853, USA.\\$^2$College of Engineering, Cornell University, Ithaca, NY 14853, USA.\\$^3$NASA Ames Research Center, Moffett Field, CA 94035, USA}

\begin{abstract}

We have conducted an intensive search using the Cassini~ISS narrow-angle camera to identify any material that may orbit Rhea.  Our results contradict an earlier and surprising inference that Rhea, the second-largest moon of Saturn, possesses a system of narrow rings embedded in a broad circum-satellite disk or cloud \citep{Jones08}.  

\textbf{Keywords:}  Rhea, Rhea rings

\textbf{Index terms:} 2756 Magnetospheric Physics: Planetary magnetospheres (5443, 5737, 6033); 5465 Planetary Sciences: Solid Surface Planets: Rings and dust; 6265 Planetary Sciences: Solar System Objects: Planetary rings; 6280 Planetary Sciences: Solar System Objects: Saturnian satellites
\end{abstract}

\section{Introduction}
During several flybys of Rhea, charged-particle detectors aboard the Cassini spacecraft recorded sharp drops in measured electrons, symmetric about the moon \citep{Jones08}, which have been interpreted as the signatures of three narrow rings, tens of~km in radial width, surrounding the moon at distances between 2~to 3~R$_{\mathrm{R}}$ from the center of the moon (Rhea's radius is 1~R$_{\mathrm{R}} = 764$~km).  In addition, a gradual decrease in charged particles was detected out to 7~R$_{\mathrm{R}}$, nearly filling the moon's Hill sphere\fn{The Hill sphere of an object is the region of its gravitational dominance.  It extends approximately out to the Hill radius $r_H = a (m/3M_S)^{1/3}$, where $m$ and $a$ are the moon's mass and orbital semimajor axis, and $M_S$ is the mass of Saturn.  Rhea's Hill radius is 5800~km = 7.6~R$_{\mathrm{R}}$.} \citep{Jones08}, which was interpreted as a broad cloud surrounding the moon.  Similar features have not been detected in the charged-particle environments around other Saturnian moons, particularly Tethys.  Charged-particle absorptions have previously been used to infer the presence of Saturn's G~ring \citep{VanAllen83} and Methone ring arc \citep{Roussos08}, as well as Jupiter's ring \citep{FMM75,BurnsChapter04}.  To date, no moon has been confirmed to possess a ring system, nor indeed has any solid body, and accordingly this interpretation of the Rhea features attracted great interest. 

\begin{figure}[!t]
\begin{center}
\includegraphics[width=16cm]{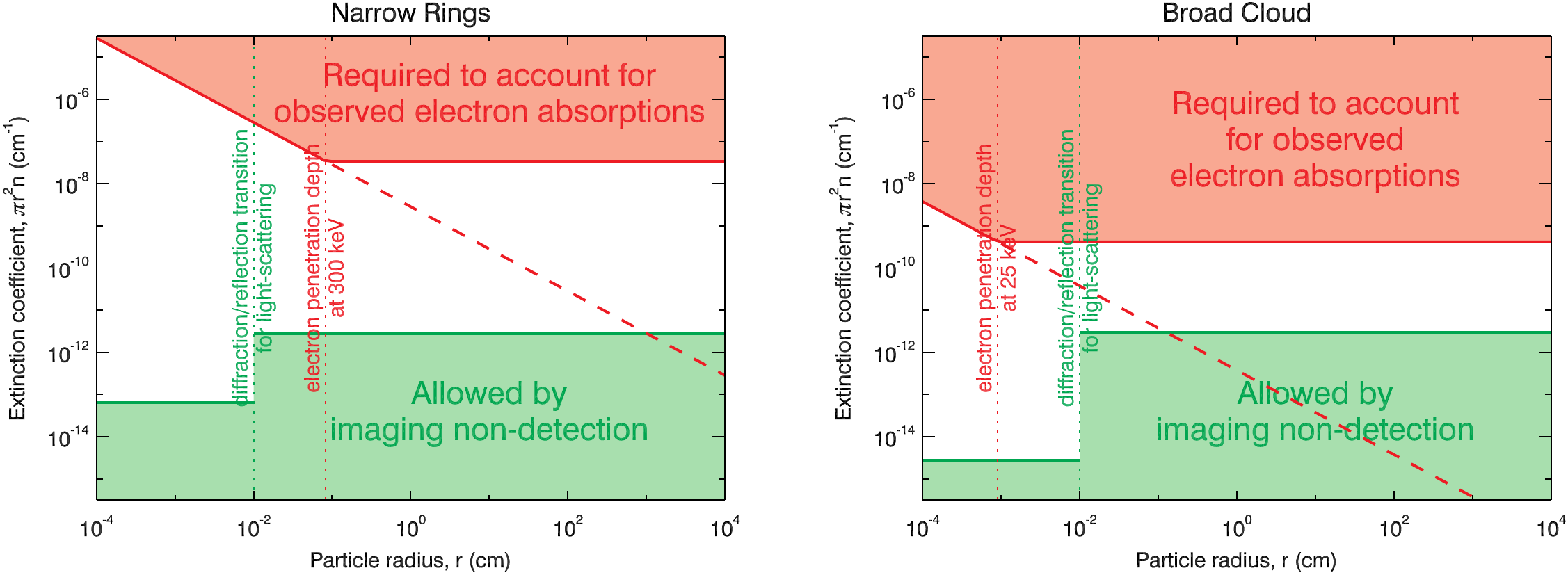}
\caption{Comparison of the radius $r$ and number density $n$ of particles, expressed in terms of the extinction coefficient $\pi r^2 n$, from observed charged-particle absorptions \citep{Jones08} and our imaging non-detection (red and green, respectively).  Dashed lines indicate requirements previously claimed \citep{Jones08} for particle sizes larger than the electron penetration depth.  For narrow rings, even allowing the latter claim, the combined observations require particles larger than 8~m in radius, indicating an unrealistic lack of smaller particles.  Once allowance is made for the role of the electron absorption length (horizontal lower boundary to red area), our imaging non-detection rules out absorption by solid material as the cause of the observed charged-particle absorptions for both narrow rings and a broad cloud.  
\label{rhearpx_fig}}
\end{center}
\end{figure}

To seek these structures, we obtained images of Rhea's equatorial plane, off the limb of the moon, using the Cassini Imaging Science Subsystem (ISS) Narrow-Angle Camera \citep{PorcoSSR04}.  Our results, described below, are summarized in Figure~\ref{rhearpx_fig}.  We conclude that both narrow rings and a broad cloud or disk around Rhea are ruled out by our observations as the cause of the observed electron absorptions.  Any narrow rings around Rhea able to account for the electron absorptions must be four orders of magnitude larger, in terms of the extinction coefficient (or total particle cross-section area per unit path length, expressed as $\pi r^2 n$ for characteristic particle radius $r$ and number density $n$), than what our observations can exclude.  Similarly, any broad cloud around Rhea must be two orders of magnitude larger in $\pi r^2 n$ than our observations allow in order to explain the electron absorptions. 

Section~\ref{Data} describes our data, Section~\ref{Analysis} our analysis, and Section~\ref{Discussion} contains further discussion and our conclusions.

\section{Data \label{Data}}

\renewcommand{\thetable}{S\arabic{table}}

\begin{table*}[!t]
\vspace{-0.7cm}
\caption{Observing information and measured RMS~$I/F$ for individual low-phase images used in this paper. \label{image_table_loph}}
\begin{scriptsize}
\begin{tabular} { c c c c c c c c c c }
\hline
\hline
 & & & Incidence & Emission & Phase & & Ring Vertical & \\
Orbit & Image Identifier & Date & Angle$^a$ & Angle$^a$ & Angle & Range (R$_\mathrm{S}$) & Width$^b$ (pixels) & RMS~$I/F$ ($10^{-6}$) \\
\hline
072 & N1592502480 & 2008-170 &  96.5$^\circ$ &  90.9$^\circ$ &   9.0$^\circ$ &  19.7 &  10.6 &   8.4 \\
072 & N1592502558 & 2008-170 &  96.5$^\circ$ &  90.9$^\circ$ &   9.0$^\circ$ &  19.7 &  10.4 &   8.2 \\
072 & N1592502636 & 2008-170 &  96.5$^\circ$ &  90.9$^\circ$ &   9.0$^\circ$ &  19.7 &  10.3 &   7.9 \\
072 & N1592502714 & 2008-170 &  96.5$^\circ$ &  90.9$^\circ$ &   8.9$^\circ$ &  19.7 &  10.2 &   7.7 \\
072 & N1592502792 & 2008-170 &  96.5$^\circ$ &  90.9$^\circ$ &   8.9$^\circ$ &  19.7 &  10.1 &   7.6 \\
072 & N1592502870 & 2008-170 &  96.5$^\circ$ &  90.8$^\circ$ &   8.9$^\circ$ &  19.7 &   9.9 &   7.5 \\
\hline
072 & N1592503520 & 2008-170 &  96.5$^\circ$ &  91.0$^\circ$ &   8.4$^\circ$ &  19.7 &   8.9 &   9.5 \\
072 & N1592503598 & 2008-170 &  96.5$^\circ$ &  91.0$^\circ$ &   8.4$^\circ$ &  19.7 &   8.8 &   9.6 \\
072 & N1592503676 & 2008-170 &  96.5$^\circ$ &  90.9$^\circ$ &   8.4$^\circ$ &  19.7 &   8.6 &   9.6 \\
072 & N1592503754 & 2008-170 &  96.5$^\circ$ &  90.9$^\circ$ &   8.4$^\circ$ &  19.7 &   8.5 &   9.6 \\
072 & N1592503832 & 2008-170 &  96.5$^\circ$ &  90.9$^\circ$ &   8.4$^\circ$ &  19.7 &   8.4 &   9.6 \\
072 & N1592503910 & 2008-170 &  96.5$^\circ$ &  90.9$^\circ$ &   8.4$^\circ$ &  19.7 &   8.2 &   9.6 \\
\hline
087 & N1601901501 & 2008-279 &  94.8$^\circ$ &  89.5$^\circ$ &  12.6$^\circ$ &  19.8 &   5.2 &   7.2 \\
087 & N1601901579 & 2008-279 &  94.8$^\circ$ &  89.5$^\circ$ &  12.6$^\circ$ &  19.8 &   5.5 &   7.2 \\
087 & N1601901657 & 2008-279 &  94.8$^\circ$ &  89.5$^\circ$ &  12.7$^\circ$ &  19.8 &   5.8 &   7.2 \\
087 & N1601901735 & 2008-279 &  94.8$^\circ$ &  89.5$^\circ$ &  12.7$^\circ$ &  19.8 &   6.0 &   7.0 \\
087 & N1601901813 & 2008-279 &  94.8$^\circ$ &  89.4$^\circ$ &  12.8$^\circ$ &  19.8 &   6.3 &   7.1 \\
087 & N1601901891 & 2008-279 &  94.8$^\circ$ &  89.4$^\circ$ &  12.8$^\circ$ &  19.8 &   6.6 &   7.1 \\
087 & N1601901969 & 2008-279 &  94.8$^\circ$ &  89.4$^\circ$ &  12.9$^\circ$ &  19.8 &   6.8 &   7.1 \\
087 & N1601902047 & 2008-279 &  94.8$^\circ$ &  89.4$^\circ$ &  12.9$^\circ$ &  19.8 &   7.1 &   7.0 \\
087 & N1601902125 & 2008-279 &  94.8$^\circ$ &  89.4$^\circ$ &  13.0$^\circ$ &  19.8 &   7.4 &   7.0 \\
087 & N1601902203 & 2008-279 &  94.8$^\circ$ &  89.3$^\circ$ &  13.0$^\circ$ &  19.8 &   7.6 &   7.0 \\
087 & N1601902281 & 2008-279 &  94.8$^\circ$ &  89.3$^\circ$ &  13.1$^\circ$ &  19.8 &   7.9 &   7.0 \\
087 & N1601902359 & 2008-279 &  94.8$^\circ$ &  89.3$^\circ$ &  13.1$^\circ$ &  19.8 &   8.2 &   7.0 \\
087 & N1601902437 & 2008-279 &  94.8$^\circ$ &  89.3$^\circ$ &  13.2$^\circ$ &  19.8 &   8.4 &   7.0 \\
087 & N1601902515 & 2008-279 &  94.8$^\circ$ &  89.3$^\circ$ &  13.2$^\circ$ &  19.8 &   8.7 &   7.1 \\
\hline
104 & N1614753658 & 2009-062 &  92.5$^\circ$ &  90.4$^\circ$ &   5.3$^\circ$ &  19.7 &   5.8 &   7.8 \\
104 & N1614753736 & 2009-062 &  92.5$^\circ$ &  90.4$^\circ$ &   5.3$^\circ$ &  19.7 &   5.5 &   7.5 \\
104 & N1614753814 & 2009-062 &  92.5$^\circ$ &  90.3$^\circ$ &   5.4$^\circ$ &  19.7 &   5.3 &   7.7 \\
104 & N1614753892 & 2009-062 &  92.5$^\circ$ &  90.3$^\circ$ &   5.4$^\circ$ &  19.7 &   5.0 &   7.4 \\
104 & N1614753970 & 2009-062 &  92.5$^\circ$ &  90.3$^\circ$ &   5.4$^\circ$ &  19.7 &   4.8 &   7.4 \\
104 & N1614754048 & 2009-062 &  92.5$^\circ$ &  90.3$^\circ$ &   5.4$^\circ$ &  19.7 &   4.5 &   7.4 \\
104 & N1614754126 & 2009-062 &  92.5$^\circ$ &  90.2$^\circ$ &   5.4$^\circ$ &  19.7 &   4.2 &   7.5 \\
104 & N1614754204 & 2009-062 &  92.5$^\circ$ &  90.2$^\circ$ &   5.5$^\circ$ &  19.7 &   4.0 &   7.5 \\
\hline
104 & N1614757423 & 2009-062 &  92.5$^\circ$ &  89.4$^\circ$ &   6.7$^\circ$ &  19.7 &   7.5 &  11.5 \\
104 & N1614757501 & 2009-062 &  92.5$^\circ$ &  89.4$^\circ$ &   6.8$^\circ$ &  19.7 &   7.8 &  11.5 \\
104 & N1614757579 & 2009-062 &  92.5$^\circ$ &  89.4$^\circ$ &   6.8$^\circ$ &  19.7 &   8.1 &  11.6 \\
104 & N1614757657 & 2009-062 &  92.5$^\circ$ &  89.3$^\circ$ &   6.8$^\circ$ &  19.7 &   8.4 &  11.6 \\
104 & N1614757735 & 2009-062 &  92.5$^\circ$ &  89.3$^\circ$ &   6.8$^\circ$ &  19.7 &   8.7 &  11.8 \\
104 & N1614757813 & 2009-062 &  92.5$^\circ$ &  89.3$^\circ$ &   6.9$^\circ$ &  19.7 &   9.0 &  11.7 \\
104 & N1614757891 & 2009-062 &  92.5$^\circ$ &  89.2$^\circ$ &   6.9$^\circ$ &  19.7 &   9.3 &  11.6 \\
104 & N1614757969 & 2009-062 &  92.5$^\circ$ &  89.2$^\circ$ &   6.9$^\circ$ &  19.7 &   9.6 &  11.6 \\
\hline
109 & N1619433032 & 2009-116 &  91.6$^\circ$ &  90.8$^\circ$ &   4.5$^\circ$ &  18.6 &  10.8 &   8.1 \\
109 & N1619433110 & 2009-116 &  91.6$^\circ$ &  90.8$^\circ$ &   4.5$^\circ$ &  18.6 &  10.5 &   8.2 \\
109 & N1619433188 & 2009-116 &  91.6$^\circ$ &  90.8$^\circ$ &   4.5$^\circ$ &  18.6 &  10.1 &   8.2 \\
109 & N1619433266 & 2009-116 &  91.6$^\circ$ &  90.8$^\circ$ &   4.5$^\circ$ &  18.6 &   9.8 &   8.2 \\
109 & N1619433344 & 2009-116 &  91.6$^\circ$ &  90.7$^\circ$ &   4.5$^\circ$ &  18.6 &   9.5 &   8.1 \\
109 & N1619433422 & 2009-116 &  91.6$^\circ$ &  90.7$^\circ$ &   4.5$^\circ$ &  18.6 &   9.2 &   8.1 \\
109 & N1619433500 & 2009-116 &  91.6$^\circ$ &  90.7$^\circ$ &   4.5$^\circ$ &  18.6 &   8.9 &   8.0 \\
109 & N1619433578 & 2009-116 &  91.6$^\circ$ &  90.6$^\circ$ &   4.5$^\circ$ &  18.6 &   8.6 &   8.1 \\
109 & N1619433656 & 2009-116 &  91.6$^\circ$ &  90.6$^\circ$ &   4.6$^\circ$ &  18.6 &   8.3 &   8.1 \\
109 & N1619433734 & 2009-116 &  91.6$^\circ$ &  90.6$^\circ$ &   4.6$^\circ$ &  18.6 &   8.0 &   8.3 \\
109 & N1619433812 & 2009-116 &  91.6$^\circ$ &  90.6$^\circ$ &   4.6$^\circ$ &  18.6 &   7.6 &   8.0 \\
109 & N1619433890 & 2009-116 &  91.6$^\circ$ &  90.5$^\circ$ &   4.6$^\circ$ &  18.6 &   7.3 &   8.1 \\
109 & N1619433968 & 2009-116 &  91.6$^\circ$ &  90.5$^\circ$ &   4.6$^\circ$ &  18.6 &   7.0 &   8.0 \\
109 & N1619434046 & 2009-116 &  91.6$^\circ$ &  90.5$^\circ$ &   4.6$^\circ$ &  18.6 &   6.7 &   7.9 \\
109 & N1619434124 & 2009-116 &  91.6$^\circ$ &  90.4$^\circ$ &   4.6$^\circ$ &  18.6 &   6.4 &   8.1 \\
109 & N1619434202 & 2009-116 &  91.6$^\circ$ &  90.4$^\circ$ &   4.7$^\circ$ &  18.6 &   6.0 &   8.1 \\
109 & N1619434280 & 2009-116 &  91.6$^\circ$ &  90.4$^\circ$ &   4.7$^\circ$ &  18.6 &   5.7 &   8.0 \\
\hline
\end{tabular}
\begin{flushleft}
\vspace{-0.1in}
$^a$ Measured from the direction of Rhea's north pole (normal to putative ring-plane), so that angles $>90^\circ$ denote the southern hemisphere. Note that Ring Opening Angle $= | 90^\circ - $ Emission Angle$|$. \\
$^b$ Vertical separation between near arm and far arm as projected onto the image plane. \\
\end{flushleft}
\end{scriptsize}
\end{table*}

\begin{table}[!t]
\vspace{1cm}
\caption{Observing information and measured RMS~$I/F$ for individual high-phase images used in this paper. \label{image_table_hiph}}
\begin{scriptsize}
\begin{tabular} { c c c c c c c c c c }
\hline
\hline
 & & & Incidence & Emission & Phase & & Ring Vertical & \\
Orbit & Image Identifier & Date & Angle$^a$ & Angle$^a$ & Angle & Range (R$_\mathrm{S}$) & Width (pixels) & RMS~$I/F$ ($10^{-6}$) \\
\hline
100 & N1610617989 & 2009-014 &  93.2$^\circ$ &  89.2$^\circ$ & 153.5$^\circ$ &   9.1 &  11.0 &  0.50 \\
100 & N1610618547 & 2009-014 &  93.2$^\circ$ &  89.7$^\circ$ & 153.2$^\circ$ &   9.1 &   4.6 &  0.45 \\
100 & N1610619105 & 2009-014 &  93.2$^\circ$ &  89.9$^\circ$ & 153.9$^\circ$ &   9.1 &   2.0 &  0.44 \\
100 & N1610619663 & 2009-014 &  93.2$^\circ$ &  90.6$^\circ$ & 152.4$^\circ$ &   9.1 &   8.6 &  0.42 \\
100 & N1610619965 & 2009-014 &  93.2$^\circ$ &  90.9$^\circ$ & 152.1$^\circ$ &   9.1 &  13.7 &  0.71 \\
100 & N1610620267 & 2009-014 &  93.2$^\circ$ &  91.2$^\circ$ & 151.8$^\circ$ &   9.1 &  17.4 &  0.47 \\
\hline
\end{tabular}
\begin{flushleft}
\vspace{-0.1in}
$^a$ Measured from the direction of Rhea's north pole (normal to putative ring-plane), so that angles $>90^\circ$ denote the southern hemisphere. \\
Note that Ring Opening Angle $= | 90^\circ - $ Emission Angle$|$. \\
\end{flushleft}
\end{scriptsize}
\end{table}

\begin{figure}[!t]
\begin{center}
\vspace{1cm}
\includegraphics[width=16cm]{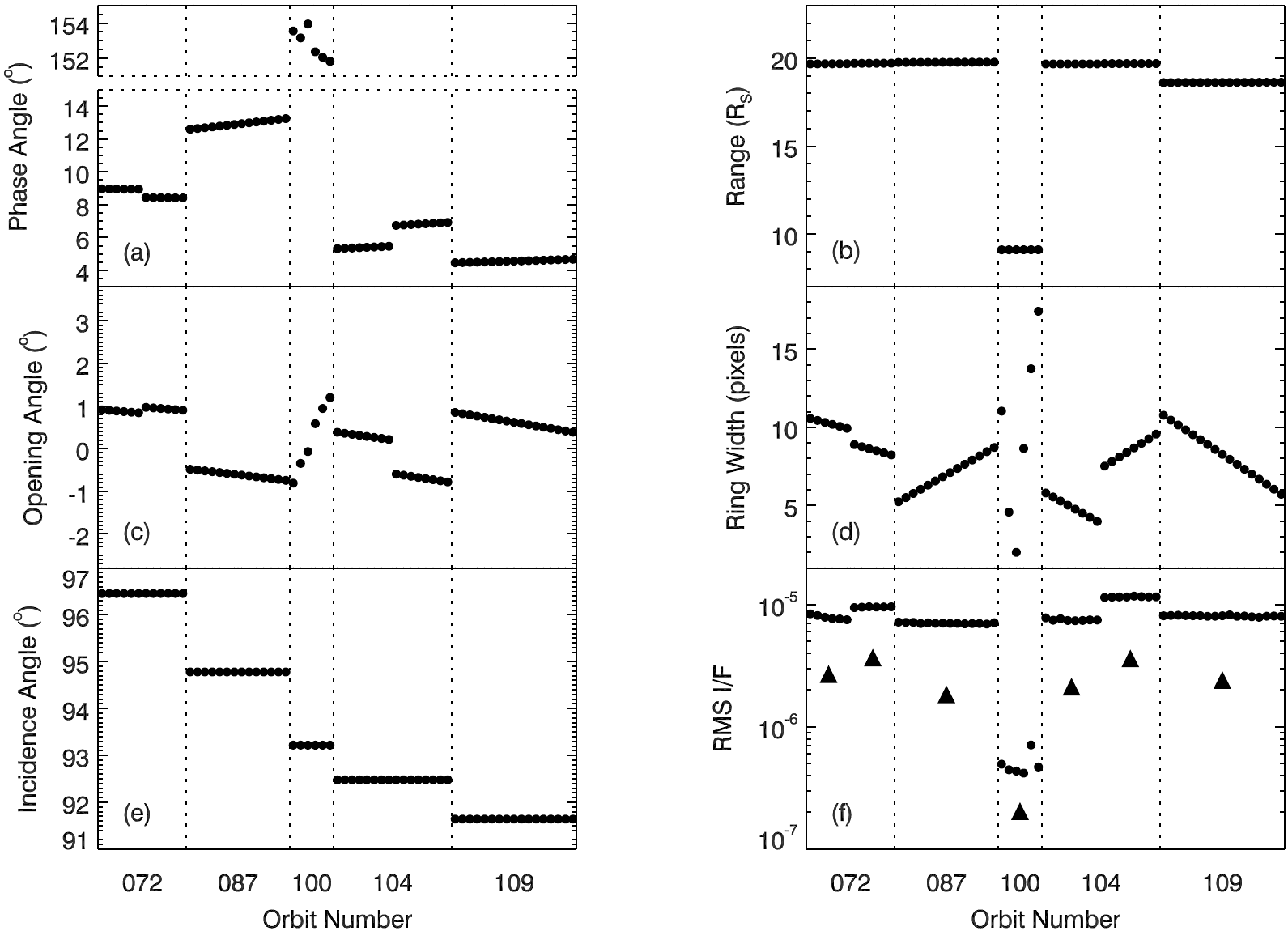}
\caption{Observing information and measured RMS~$I/F$ for images used in this paper.  Circles indicate individual images while, in panel (f), triangles indicate co-added images. \label{rhea_reproject_rms}}
\end{center}
\end{figure}

We obtained 65~clear-filter images (Tables~\ref{image_table_loph} and~\ref{image_table_hiph}, Figure~\ref{rhea_reproject_rms}) of Rhea's equatorial plane, off the limb of the moon, using the Cassini Imaging Science Subsystem (ISS) Narrow-Angle Camera \citep{PorcoSSR04}.  The 6~images at high phase angles, most sensitive to diffraction by micron-sized dust particles, were 220-second or 460-second exposures.  The 59~images at low phase angles, most sensitive to reflection by particles that are large compared to the wavelength of light, were limited to 18-second exposures due to glare from nearby Rhea at near-full phase.  All images were obtained at small values of the sub-spacecraft latitude on Rhea (i.e., ring opening angle), so that the equatorial plane was viewed nearly edge-on.  The camera was oriented so that known image artifacts were at an angle to the edge-on equatorial plane.  The images were calibrated with the latest version (v3.6) of the standard CISSCAL package \citep{PorcoSSR04}. 

\section{Analysis \label{Analysis}}

\subsection{Searching for narrow rings}

\renewcommand{\thefigure}{S\arabic{figure}}
\setcounter{figure}{0}

\begin{figure*}[!t]
\begin{center}
\includegraphics[width=7.5cm]{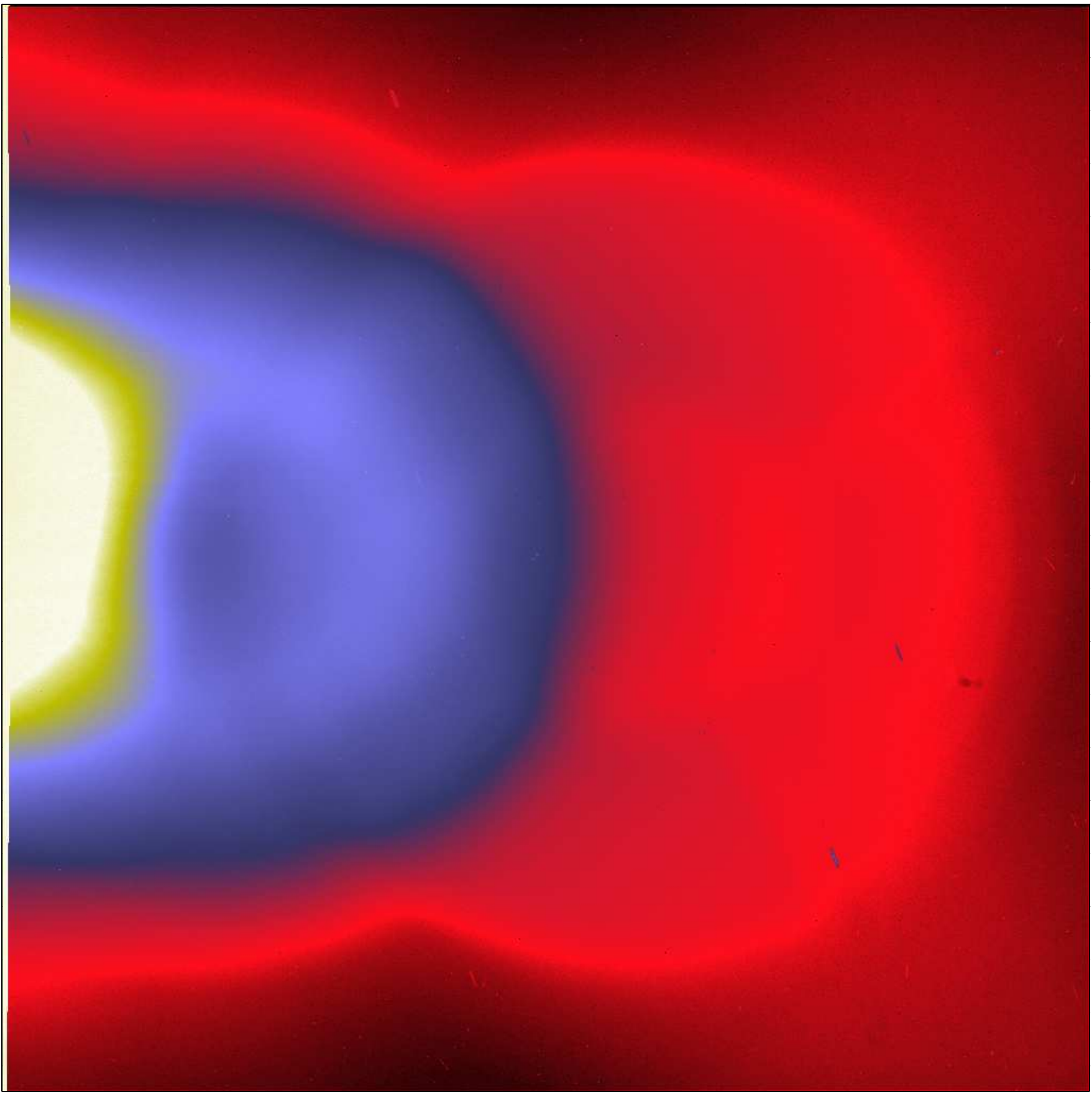}
\hspace{0.5cm}
\vspace{0.5cm}
\includegraphics[width=7.5cm]{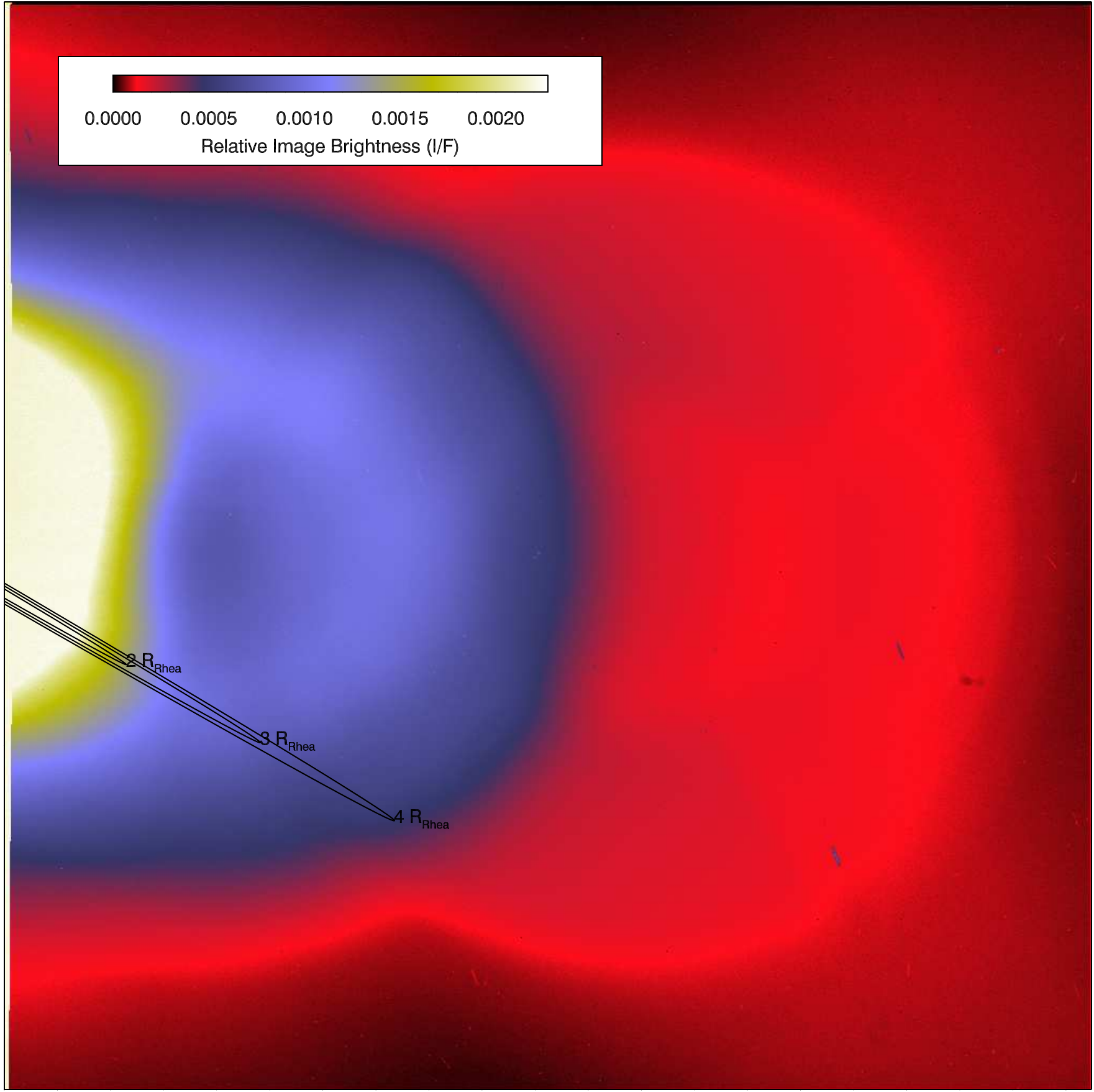}
\caption{Image N1592503520, displayed with a color stretch (see scale bar).  The two panels are identical except for contours in the right-hand panel, showing distances of 2, 3, and 4 R$_{\mathrm{R}}$ within Rhea's equatorial plane.  Saturated pixels within 1.7~R$_{\mathrm{R}}$ are due to scattered light from Rhea.  The image is characterized by irregularly-structured scattered light due to the presence of a bright light source (namely, Rhea) just outside the field of view. \label{scatlight_loph}}
\end{center}
\end{figure*}

\begin{figure*}[!t]
\begin{center}
\vspace{0.65cm}
\includegraphics[width=16cm]{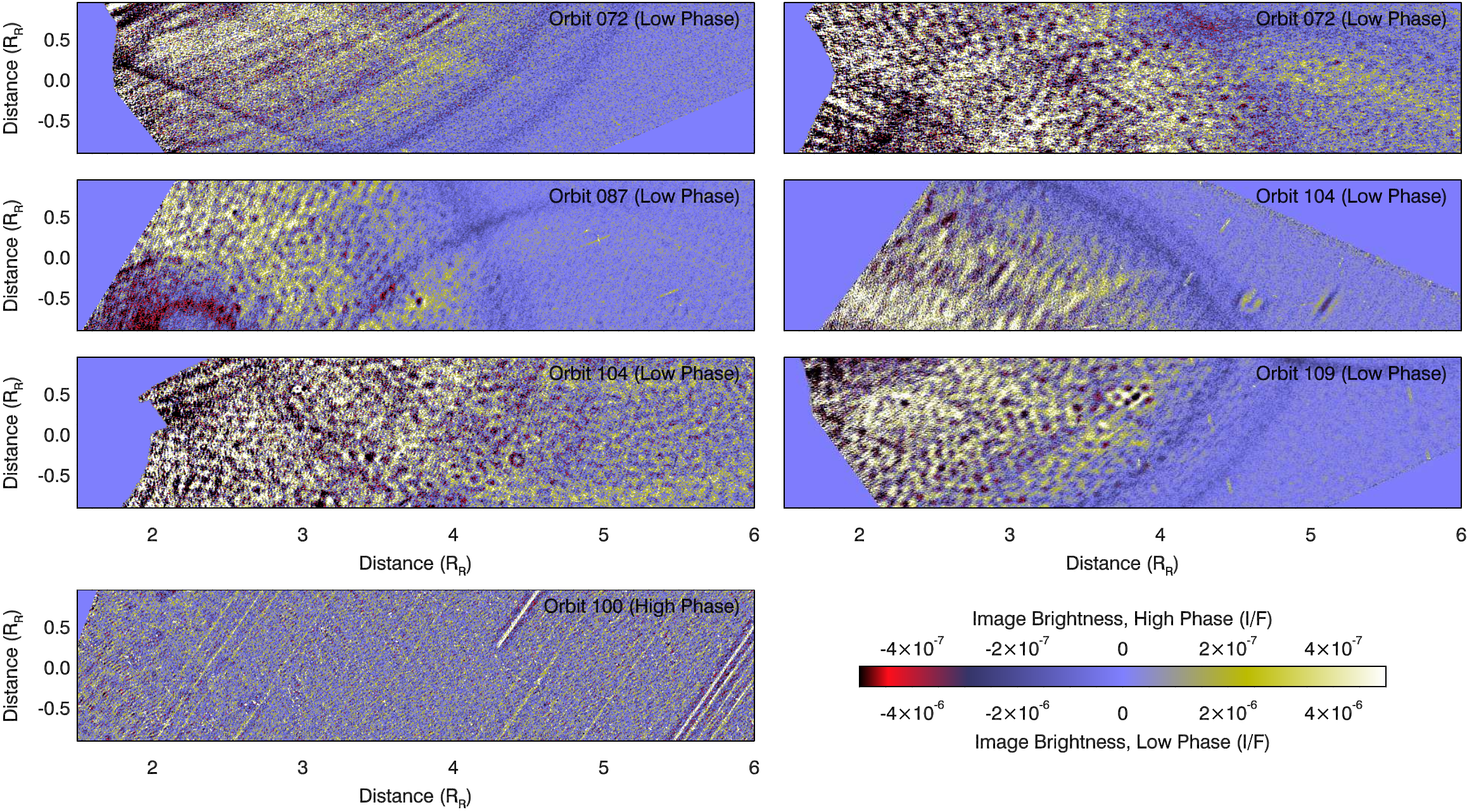}
\caption{Reduced co-added images.  
Only images taken at nearly the same time and geometry are co-added. Input images are grouped as in Tables~\ref{image_table_loph} and~\ref{image_table_hiph}.  Distances for these reprojected images are from Rhea's rotational axis (horizontal) and midplane (vertical).  Visible features in these images are either image artifacts or star streaks.  A narrow ring, if it exists, should be aligned horizontally near the middle-left of each panel (see Figure~\ref{rhea_reproject_fake_ring}).  \label{rhea_reproject_grp_display}}
\end{center}
\end{figure*}

Before co-adding images together, we first processed each image with a low-pass sigma filter (which uses a 15-pixel boxcar to replace with the boxcar mean any pixels with values more than 3 standard deviations larger than the boxcar mean) to remove cosmic rays and hot pixels, then with a high-pass boxcar filter with a 21-pixel boxcar length to remove the large-scale scattered light patterns.  The latter step was used due to the difficulty of modeling the highly irregular scattered light patterns, especially in low-phase images, caused by a bright light source (namely, Rhea) just outside the field of view (Figure~\ref{scatlight_loph} shows a typical image).  To avoid edge effects, we zeroed out any pixels in the filtered image within one boxcar length of the image edge.  We then co-added images by rotating them to orient Rhea's equatorial plane in the horizontal direction and stretching them to a common radial scale.  For rotating and stretching, we oversampled by a factor of $\sim 2$ so as not to lose information.  Because the ring opening angle is uniformly very low, we did not change the vertical scale of the images, as that would cause severe oversampling in some cases.  To reduce the effects of changing geometry, we only directly co-added images that were taken at the same time (Table~\ref{image_table_coadd}, Figure~\ref{rhea_reproject_grp_display}). 

\begin{table}[!t]
\caption{Measured RMS~$I/F$ for co-added images. \label{image_table_coadd}}
\begin{scriptsize}
\begin{tabular} { c c c c c c c }
\hline
\hline
 & & & \# of & Phase & \\
Orbit & Image Identifier & Date & Images & Angle & RMS~$I/F$ ($10^{-6}$) \\
\hline
072 & N1592502480 -- 02870 & 2008-170 &  6 &   9.0$^\circ$ &   2.7 \\
072 & N1592503520 -- 03910 & 2008-170 &  6 &   8.4$^\circ$ &   3.7 \\
087 & N1601901501 -- 02515 & 2008-279 & 14 &  12.9$^\circ$ &   1.8 \\
100 & N1610617989 -- 20267 & 2009-014 &  6 & 152.8$^\circ$ &  0.20 \\
104 & N1614753658 -- 54204 & 2009-062 &  8 &   5.4$^\circ$ &   2.1 \\
104 & N1614757423 -- 57969 & 2009-062 &  8 &   6.8$^\circ$ &   3.6 \\
109 & N1619433032 -- 34280 & 2009-116 & 17 &   4.6$^\circ$ &   2.4 \\
\hline
\end{tabular}
\end{scriptsize}
\end{table}

\begin{figure*}[!t]
\begin{center}
\includegraphics[width=16cm]{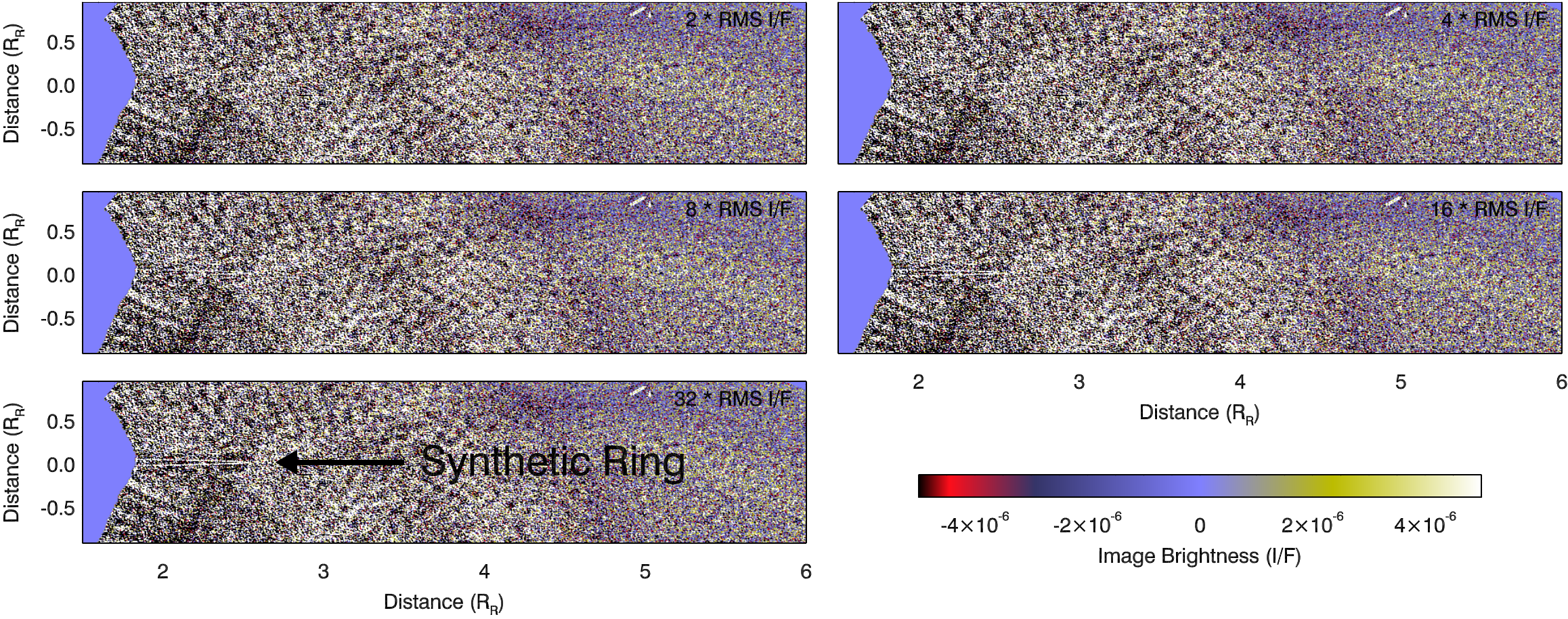}
\caption{Reduced image N1592503520 with synthetic ring added (before image reduction) at amplitudes ranging from 2 to 32 times the image's RMS~$I/F$.  The synthetic ring, which is very clearly visible and labeled in the bottom-left panel, is seen at amplitudes $\gtrsim 8$ times the RMS~$I/F$.  
\label{rhea_reproject_fake_ring}}
\end{center}
\end{figure*}

We did not find any indication of rings in Rhea's equatorial plane, either in individual or in co-added images (Figure~\ref{rhea_reproject_grp_display}).  Features that do appear in the images are either image artifacts or star streaks; none are horizontally aligned as is Rhea's equatorial plane.  To quantify our non-detection, we calculated the root-mean-square brightness (in units of $I/F$)\fn{The dimensionless quantity $I/F$ is a measure of reflected or transmitted brightness, normalized by the incident solar flux such that a perfect Lambertian surface will have $I/F = 1$.} of each individual image, and of each co-added image, in the region between 2~and 3~R$_{\mathrm{R}}$ (Figure~\ref{rhea_reproject_rms}f).  We added a synthetic ring to an unreduced image and then performed the above-described image-reduction process on it (Figure~\ref{rhea_reproject_fake_ring}).  Our synthetic ring had a semimajor axis of 2.5~R$_{\mathrm{R}}$ and a radial width of 30~km; since it was only resolved at the ansa, the apparent brightness elsewhere was reduced\fn{Our synthetic ring does not explicitly have any vertical thickness; however, adding vertical thickness to the model would have the effect of decreasing the $I/F$ in each pixel necessary to see a ring of given surface brightness (since the arms of the ring would be closer to being resolved) while simultaneously increasing the path-length factor $\mu$ (which cannot be less than the ring's vertical thickness divided by its radial width).  Overall, consideration of larger values of vertical thickness would only result in inserting factors of order unity into our calculations.}.  Since the synthetic ring becomes visible when its brightness is approximately eight times the RMS~$I/F$, we take that as the level of the non-detection. 

We conclude that our observations rule out any narrow rings around Rhea within the following constraints:  dusty rings are limited by the high-phase images to have a brightness $I/F < 1.6 \times 10^{-6}$ in the region $r > 1.4$~R$_{\mathrm{R}}$, while rings made of macroscopic particles are limited by the low-phase images to have $I/F < 1.6 \times 10^{-5}$ in the region $r > 1.7$~R$_{\mathrm{R}}$.  To convert these values to normal optical depth, we use the expression \citep{Cuzzi84}

\begin{equation}
\label{ioverf2tau}
\tau = \frac{4 \mu I/F}{\varpi_0 P(\alpha)} , 
\end{equation}

\noindent where the sine of the ring opening angle is $\mu = 0.015$ for a typical opening angle of 0.85$^\circ$ (see Figure~\ref{rhea_reproject_rms}c).  For micron-sized dust, we use a single-scattering albedo $\varpi_0 = 0.5$ and phase function $P(\alpha) \approx 3$, the latter value being from a Mie phase function calculated for water-ice spheres with a power-law size distribution index $q=3$ \citep{PC80,Show92} at phase angle $\alpha \approx 153^\circ$.  For large particles, we use a single-scattering albedo $\varpi_0 = 0.07$ (derived \citep{Cuzzi85} from the 5\% resolved-surface albedo of the dark side of Iapetus \citep{JaumannChapter09}, which we use for the sake of argument though it is surely an under-estimate) and $P(\alpha) \approx 5$ from a Callisto phase function \citep{Cuzzi84} at $\alpha \approx 6^\circ$.  Thus, we derive from our non-detection the following limits on the normal optical depth of any narrow ring around Rhea: 

\begin{equation}
\label{taulimits}
\tau_{\mathrm{dust}} < 6 \times 10^{-8}, \hspace{1.3cm}
\tau_{\mathrm{large}} < 3 \times 10^{-6} .
\end{equation}

In the case of narrow rings, the normal optical depth can be expressed as $\tau = \pi r^2 z n$ for characteristic particle size $r$ and number density $n$.  The ring's vertical thickness $z$ cannot be greater than its radial width without invoking an implausibly non-isotropic distribution of random velocities; in fact all known narrow rings are significantly flattened \citep{MWC07}.  As the radial width has been measured \citep{Jones08} to be ``tens of km'', so we will take the radial width to be $\Delta r \sim 30$~km and the vertical thickness\fn{The value we choose for the vertical thickness turns out to have an equal effect on nearly all the quantities calculated in this work, and thus to be unimportant to our final conclusions.} to be $z \sim 10$~km.  We can then express our detection limits for narrow rings in terms of the extinction coefficient, $\pi r^2 n = \tau / z$, 

\begin{eqnarray}
\hspace{2cm} (\pi r^2 n)_{\mathrm{dust}} < 6 \times 10^{-14} \hspace{0.1cm} \mathrm{cm}^{-1} , \nonumber \\
(\pi r^2 n)_{\mathrm{large}} < 3 \times 10^{-12} \hspace{0.1cm} \mathrm{cm}^{-1} .
\label{r2nlimits}
\end{eqnarray}

The absorption of electrons is proportional to the mass available to interact with the electrons, which is the mass of the ring particles as long as their sizes are small compared to the electron penetration depth, while interaction with light is proportional to the combined surface area of the ring particles.  In this limit, we can express our detection limit in terms of a minimum particle size that will remain consistent with our non-detection while incorporating enough mass behind a unit surface area to account for the observed electron absorption.  The latter constraint has been stated \citep{Jones08} as $r^3 n = 3 \times 10^{-11}$, which we multiply by a scalar to give the total particle volume along the electron path length $(4/3) \pi r^3 n = 1.3 \times 10^{-10}$.  
However, the calculations yielding that number assumed an implausibly large ring vertical thickness (i.e., the path length for the observed electrons, which have a low pitch angle $\alpha \sim 10^\circ$) of $z = 210$~km (see previous paragraph), so we multiply by a factor of $210 / 10 = 21$ to obtain $(4/3) \pi r^3 n = 3 \times 10^{-9}$.  
Dividing this by our detection limits in Equation~\ref{r2nlimits} and solving for the limiting characteristic particle size $r$, we find

\begin{equation}
\label{rlimits}
r_{\mathrm{dust}} > 300 \hspace{0.1cm} \mathrm{m} , \hspace{1.3cm}
r_{\mathrm{large}} > 8 \hspace{0.1cm} \mathrm{m} .
\end{equation}

\noindent The lower bound on $r_{\mathrm{dust}}$ is a contradiction (dust is, by definition, composed of $\mu$m-size particles, so that the particles are not very large compared to the wavelength of light), meaning that no dusty ring can account for the electron absorptions while remaining within our detection limits.  On the other hand, for large particles, at $r=8$~m the mass needed to explain the observed charged-particle absorptions can be made up with only $\sim 5000$~objects; with such a sparse population, it is questionable whether the charged-particle absorptions would be detected collectively rather than individually.  Moreover, no known population in the solar system has such large particles without an accompanying population of smaller particles in a size distribution, produced by collisions and erosion \citep{BurnsChapter01}; the latter would have been seen in our images.  Electromagnetic forces might be invoked to sweep away micron-sized dust, and a dense population of large objects can sweep up smaller ones as regolith (the latter process accounts for the lack of dust in Saturn's main rings \citep{CuzziChapter09}), but neither of these processes can remove meter-size or even cm-size objects in an optically thin ring.

A further contradiction is present in the above analysis, namely that the lower limit on $r_{\mathrm{large}}$ given in Equation~\ref{rlimits} is much larger than the electron penetration depth.  At such sizes, the ability of particles to block electrons becomes proportional not to their mass but to their surface area, just like their ability to interact with light.  For 300-keV electrons, the electron penetration depth\fn{See ``Stopping-power and range tables for electrons'', on the website of the National Institute of Standards and Technology, http://physics.nist.gov/PhysRefData/Star/Text/ESTAR.html} for water ice with density $\rho \sim 1$~g~cm$^{-3}$ is $d_e = 0.08$~cm.  So the total particle volume along the electron path length, derived from the measured electron absorption as discussed in the previous paragraph, must be divided by the skin depth $d_e$ to yield an extinction coefficient $\pi r^2 n = 3 \times 10^{-8}$~cm$^{-1}$.  
This is several orders of magnitude larger than the imaging detection limit for the same quantity in Equation~\ref{r2nlimits}, meaning that it is impossible for electron absorption by any narrow ring to account for the observations. 

\subsection{Searching for a broad cloud}

\begin{figure*}[!t]
\begin{center}
\includegraphics[width=7.5cm]{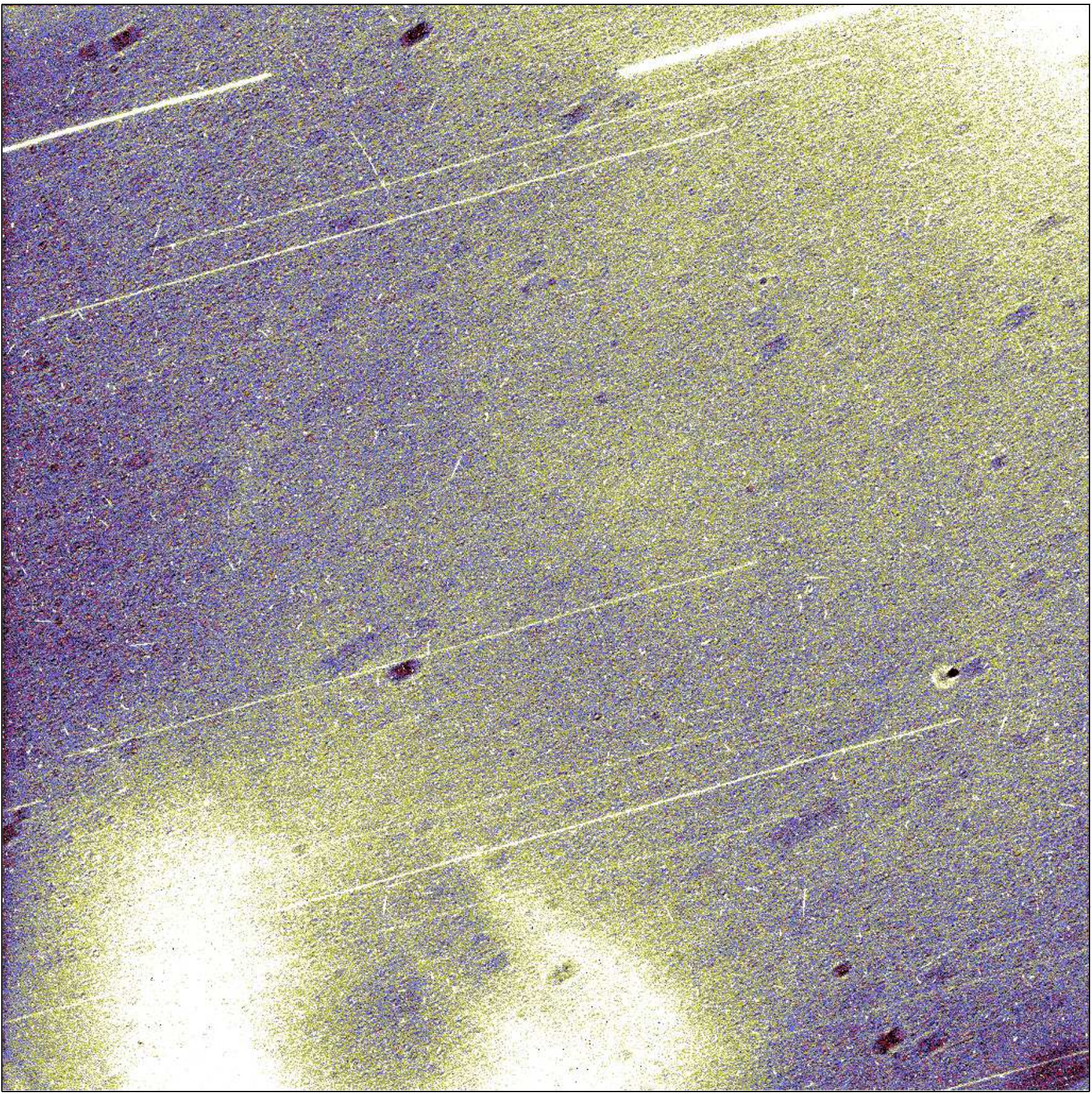}
\hspace{0.5cm}
\vspace{0.5cm}
\includegraphics[width=7.5cm]{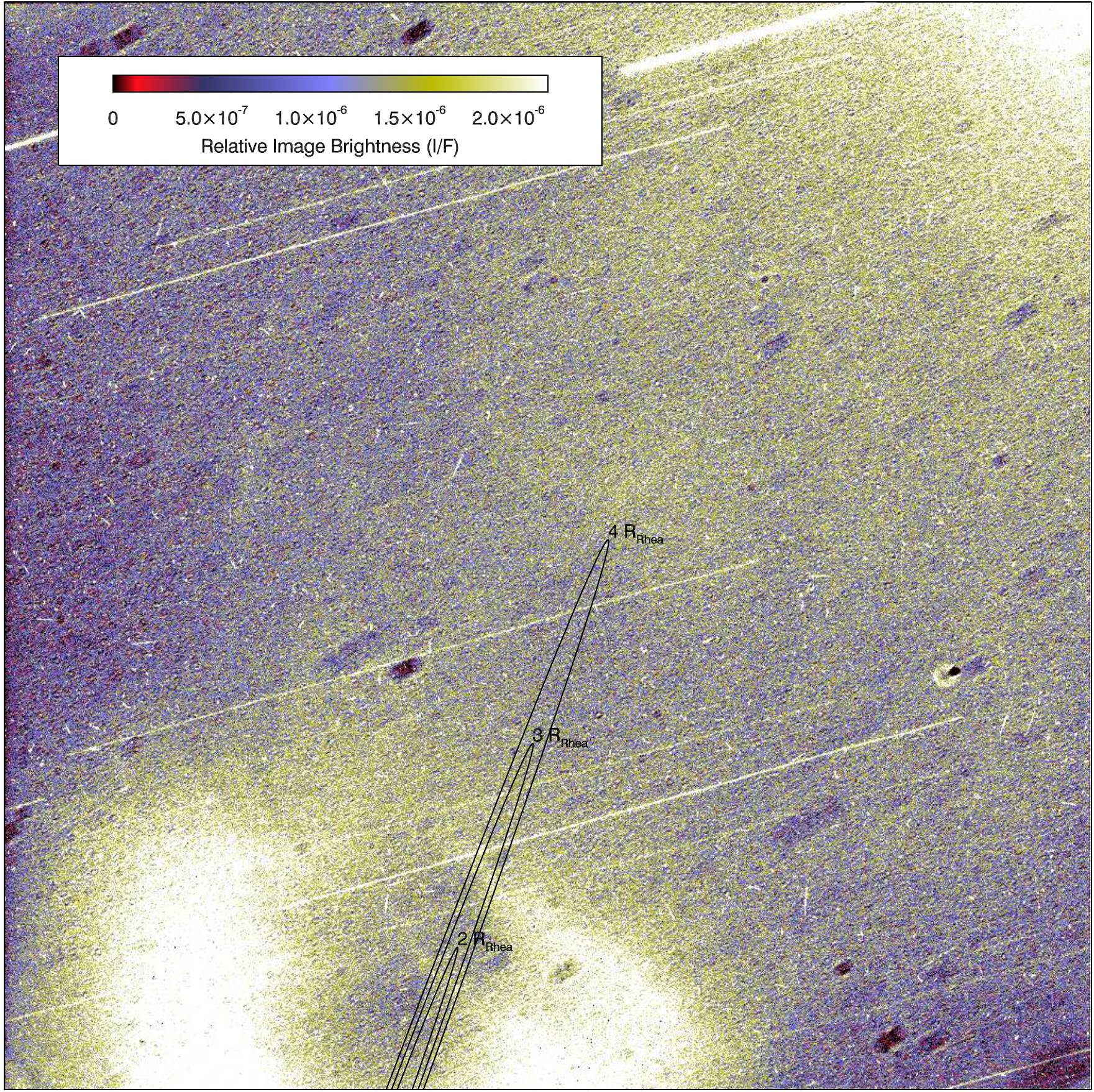}
\caption{Image N1610620267, displayed with a color stretch (see scale bar).  The two panels are identical except for contours in the right-hand panel, showing distances of 2, 3, and 4 R$_{\mathrm{R}}$ within Rhea's equatorial plane.  The long streaks trending $\sim 16^\circ$ counter-clockwise from horizontal are stars, smeared over the 220-second exposure.  The numerous localized features remain stationary in all six of the high-phase images Orbit~100 images (Table~\ref{image_table_hiph}) 
and thus must be due to imperfections in the camera; although these are not catalogued in the general calibration \citep{WestCal10}, we believe they are attributable to the unusually high phase and/or the unusually asymmetric lighting (R.~West and B.~Knowles, personal communication, 2010).  The irregularly-structured scattered light in the lower-left corner is due to the presence of a bright light source (namely, Rhea) just outside the field of view.  The scattered light emanating from the upper-right corner is due to internal reflections within the camera, but nevertheless we set our detection limit to be greater than its amplitude. \label{scatlight_hiph}}
\end{center}
\end{figure*}

Unlike the low-phase images, which were taken with the nearly-full disk of Rhea off the edge of the field of view, the high-phase images are not contaminated with a great deal of scattered light.  We did not find any indication of a broad cloud surrounding Rhea in these images (Figure~\ref{scatlight_hiph}).  Features that do appear in the images are either image artifacts or star streaks; none show any indication of Rhea-centered symmetry.  To quantify our non-detection, we use the amplitude of irregularly-structured scattered light in the region beyond 3~R$_{\mathrm{R}}$.  We conclude that our observations rule out any broad cloud around Rhea unless the brightness is less than $I/F = 10^{-6}$. 

We again use Equation~\ref{ioverf2tau} with the same input values, except for the following: for large particles we now must use the Callisto phase function at high phase, yielding $P(\alpha) \approx 0.2$ for $\alpha \approx 153^\circ$ \citep{Cuzzi84,Dones93}; for a spherical cloud we set $\mu$ (which really represents a ratio of path lengths) to unity; and, of course, we have new values of $I/F$.  We thus derive limiting optical depths for the broad cloud of 

\begin{equation}
\label{taulimits_cloud}
\tau_{\mathrm{cloud,dust}} < 3 \times 10^{-6} , \hspace{0.8cm}
\tau_{\mathrm{cloud,large}} < 3 \times 10^{-3} . 
\end{equation}

\noindent Using the simplest available model, of a uniform spherical cloud with radius 7~R$_{\mathrm{R}}$, a ray through the cloud with a closest approach at 3~R$_{\mathrm{R}}$ has a path length of $x_{\mathrm{cloud}} \sim 2 \sqrt{7^2-3^2} \sim 13$~R$_{\mathrm{R}}$. 
Thus we can write our optical depth as $\tau = \pi r^2 x_{\mathrm{cloud}} n$ and express our detection limit as 

\begin{eqnarray}
 \hspace{1.6cm} (\pi r^2 n)_{\mathrm{cloud,dust}} < 3 \times10^{-15} \hspace{0.1cm} \mathrm{cm}^{-1} , \nonumber \\
(\pi r^2 n)_{\mathrm{cloud,large}} < 3 \times10^{-12} \hspace{0.1cm} \mathrm{cm}^{-1} . 
\label{r2nlimits_cloud}
\end{eqnarray}

\noindent Combining the first of these two results, as in the previous section, with the value inferred from electron absorption for the broad cloud $r^3 n = 9 \times 10^{-14}$, we again get a contradictory limit of $r_{\mathrm{cloud,dust}} > 1$~m.  For a halo made of macroscopic particles, however, both the electron absorption and our non-detection can be accounted for as long as $r_{\mathrm{cloud,large}} > 1$~mm.  
However, we again have the problem that this lower limit on the particle size is much larger than the electron penetration depth$^\P$, which for 24-28~keV electrons is only $d_e = 9 \times 10^{-4}$~cm = 9~$\mu$m \citep{Jones08}.  So, as before, we divide the quantity $(4/3) \pi r^3 n$ by the skin depth $d_e$, yielding a new value for the observed electron absorption of $\pi r^2 n = 4 \times 10^{-10}$~cm$^{-1}$.  
As shown in Figure~\ref{rhearpx_fig}, this again is several orders of magnitude too large,\fn{On the other hand, the Cosmic Dust Analyzer (CDA), which directly samples the dust surrounding \textit{Cassini}, detected an increase in $>1$~$\mu$m particles of $\sim 5 \times 10^{-4}$~m$^{-3}$ in Rhea's vicinity \citep[Figure~3a of][]{Jones08}.  This works out to an extinction coefficient of $(\pi r^2 n)_{\mathrm{CDA}} = 2 \times 10^{-17}$~cm$^{-1}$, 
which is well below our detection threshold (Equation~\ref{r2nlimits_cloud}, Figure~\ref{rhearpx_fig}) and thus does not contradict our observations.} even for the relatively permissive limit for large particles in Equation~\ref{r2nlimits_cloud}, meaning that it is impossible for electron absorption by any broad cloud to account for the observations. 

\section{Discussion and Conclusions \label{Discussion}}
The absorption of electrons by solid particles is proportional to the particles' mass, as long as their sizes are small compared to the electron penetration depth.  However, for particles significantly larger than the electron penetration depth, only the particle surfaces are available to interact with charged particles \citep{VanAllen83,VanAllen87}.  Based on this, the range of $\pi r^2 n$ required to account for the magnetospheric observations is shown by the red areas in Figure~\ref{rhearpx_fig}.  However, in their calculations of the electron absorptions needed to explain their observations, \citet{Jones08} assumed that a particle's absorption power is proportional to its volume for all particle sizes, even those much larger than the penetration depth.  The dashed lines in Figure~\ref{rhearpx_fig} indicate what the required ring parameters would be if this assumption were valid.  In such a case, our images, which constrain the surface area of particles as they interact with photons, would not be able to rule out a broad cloud of mm-size particles around Rhea; however, our observations would only allow narrow rings made up primarily of particles larger than 8~m (see where the dashed red lines cross the green lines in Figure~\ref{rhearpx_fig}).  This is problematic because all known solar-system populations of such large objects are accompanied by even more smaller objects produced by collisions and erosion \citep{BurnsChapter01}, and the latter would have been seen in our images.  Electromagnetic forces might be invoked to sweep away micron-sized dust, and a dense population of large objects can sweep up smaller ones as regolith (the latter process accounts for the lack of dust in Saturn's main rings \citep{CuzziChapter09}), but neither of these processes can remove meter-size or even cm-size objects in an optically thin ring.

The details of interactions between charged particles and solid matter in Saturn's magnetosphere may well be more complex than has heretofore been considered.  \citet{Jones08} suggested that treating the full kinematics of the problem with Monte Carlo collision codes could resolve the difficulties discussed here and in their paper.  However, it is unlikely that corrections of this kind will bridge the gap of several orders of magnitude shown in Figure~\ref{rhearpx_fig}.

We conclude that neither narrow rings nor a broad disk or cloud around Rhea is likely consistent with the available observations.  However, the detection of highly unusual charged-particle absorptions in Rhea's vicinity is certainly real, and likely advertises some new mechanism.  Thus, we urge the magnetospheric community to find alternate explanations that do not invoke solid matter orbiting the moon. 

\begin{acknowledgments}
We thank G.~H.~Jones and E.~Roussos for helpful discussions, and B.~M.~Byington for assistance with image processing.  We acknowledge K.~Perry, B.~Wallis, the Cassini Project, and the Cassini Imaging Team for making these observations possible.  We acknowledge funding from the Cassini Project and from NASA's Cassini Data Analysis Program (NNX08AQ72G and NNX10AG67G). 
\end{acknowledgments}


\end{document}